# Electron shock drift acceleration at a low-Mach-number, low-plasma-beta quasi-perpendicular shock

Ao Guo[1,2], Quanming Lu[1,2,3*], San Lu[1,2,3], Zhongwei Yang[4], and Xinliang Gao[1,2,3]

[1]CAS Key Lab of Geospace Environment, School of Earth and Space Sciences, University of Science and Technology of China, Hefei 230026, China

[2]CAS Center for Excellence in Comparative Planetology, Hefei 230026, China

[3]Collaborative Innovation Center of Astronautical Science and Technology, Harbin, China

[4]State Key Laboratory of Space Weather, National Space Science Center, Chinese Academy of Sciences, Beijing 100190, People's Republic of China

*Corresponding Author: Quanming Lu, Email:qmlu@ustc.edu.cn

## Abstract

Shock drift acceleration plays an important role in generating high-energy electrons at quasi-perpendicular shocks, but its efficiency in low beta plasmas is questionable. In this article, we perform a two-dimensional particle-in-cell simulation of a low-Mach-number low-plasma-beta quasi-perpendicular shock, and find that the electron cyclotron drift instability is unstable at the leading edge of the shock foot, which is excited by the relative drift between the shock-reflected ions and the incident electrons. The electrostatic waves triggered by the electron cyclotron drift instability can scatter and heat the incident electrons, which facilitates them to escape from the shock's loss cone. These electrons are then reflected by the shock and energized by shock drift acceleration. In this way, the acceleration efficiency of shock drift acceleration at low-plasma-beta quasi-perpendicular shocks is highly enhanced.

# 1. Introduction

Collisionless shocks play a crucial role in accelerating charged particles in space and astrophysical plasmas. Coherent emissions from supernova remnants (Koyama et al. 1995; Ellison 2001; Reynolds 2008) and coronal mass ejections (CMEs) (Holman & Pesses 1983; Pulupa & Bale 2008) suggest efficient electron acceleration by collisionless shocks. High-energy electrons are also observed in situ at planetary bow shocks (Masters et al. 2013, 2017; Liu et al. 2019; Lindberg et al. 2024). The diffusive shock acceleration (DSA) theory (Bell 1978; Blandford & Ostriker 1978) has been applied to explain these phenomena, which successfully predicts a power-law distribution of accelerated high-energy electrons. In DSA, electrons are accelerated through crossing the shock back and forth repeatedly, which needs their gyroradii to be comparable to the shock thickness before acceleration (Balogh & Treumann 2013; Grassi et al. 2023; Lindberg et al. 2023). Therefore, a pre-acceleration is necessary for the thermal electrons to enter the DSA process.

Shock drift acceleration (SDA) has been considered as the most possible candidate for electron pre-acceleration in quasi-perpendicular shocks (Matsukiyo et al. 2011; Park et al. 2012, 2013), where the angle between the upstream magnetic field and the shock normal ($\theta_{Bn}$) is larger than 45°. In SDA, the upstream electrons are reflected by the magnetic mirror force, and then get accelerated by the convection electric field during their gradient-B drift (Wu 1984; Krauss-Varban & Wu 1989; Ball & Melrose 2001; Mann et al. 2006; Yang et al. 2009; Park et al. 2013). However, only those electrons with their perpendicular velocities in the upstream frame satisfying $v_\perp \gtrsim \frac{U_{sh}}{\cos\theta_{Bn}} \sin\theta_c$ can be reflected by the shock (where $U_{sh}$ is the shock speed, $\theta_c = \sin^{-1}\left(\sqrt{\frac{B_0}{B_\mathrm{m}}}\right)$ is the loss-cone angle of the shock, and $B_0$ and $B_\mathrm{m}$ are the magnetic field strength in the upstream and overshoot region, respectively). Therefore, the upstream electrons at a low-plasma-beta quasi-perpendicular shock are difficult to be reflected and then enter

SDA.

At high-Mach-number quasi-perpendicular shocks, Buneman instability is driven unstable because of the interaction between the upstream electrons and shock-reflected ions (Hoshino & Shimada 2002; Amano & Hoshino 2007, 2009; Matsumoto et al. 2012, 2013, 2017). Electrostatic (ES) waves induced by Buneman instability at the leading edge of the shock are observed to trap and accelerate electrons before they reach the shock ramp (Amano & Hoshino 2009; Matsumoto et al. 2012, 2013, 2017). These pre-energized electrons are more likely to be reflected back upstream later. However, Buneman instability can only be triggered when the shock Mach number is sufficiently high ($M_A \sim 30$) (Matsumoto et al. 2012).

At low-Mach-number quasi-perpendicular shocks, electron acceleration can be enhanced by shock surface ripples (Kobzar et al. 2021), as they may lead to multiple interactions between electrons and the shock front. Besides, interactions with high-frequency wave activities can significantly affect electron acceleration. At the Earth's bow shock, high-frequency coherent whistlers are observed to confine electrons within the shock ramp and provide more efficient acceleration than SDA (Oka et al. 2017; Amano et al. 2020). Upstream of the shock front, microinstabilities driven by the reflected ion beams can also produce high-frequency fluctuations. When the relative velocity between ions and electrons is slower than the electron thermal velocity, the modified two-stream instability (MTSI) becomes dominant at the shock foot (Umeda et al. 2009, 2012; Riquelme & Spitkovsky 2011). MTSI results from the interaction between ions and obliquely propagating electromagnetic whistler mode waves. ES waves are also excited on the branch of electron Bernstein mode by the ion beam, which is called electron cyclotron drift instability (ECDI) (Muschietti et al. 2006, 2013; Yang et al. 2020). However, how this instability interact with electrons and affect their acceleration remains unclear. The interplay between microinstabilities and the traditional SDA model becomes an important problem. In this article, we present findings from a two-dimensional (2D) particle-in-cell (PIC) simulation of a low-Mach-number low-plasma-beta quasi-perpendicular shock. ES waves at Debye scale are observed at the leading edge of the shock foot due to the exicitation of ECDI. These waves can efficiently trap and scatter the upstream electrons out of the shock's loss cone, which significantly increase the number of electrons entering the SDA process eventually. This process can solve the problem of insufficient electron reflection rate by shocks in low-plasma-beta environments.

## 2. Simulation Model

The simulation is carried out by using an open-source relativistic full PIC code called SMILEI (Derouillat et al. 2018). A 2D simulation box with a size $L_x = 122 d_{i0}$, $L_y = 10 d_{i0}$ is employed in the $x - y$ plane, where $d_{i0} = c/\omega_{pi0}$ is the ion inertial length in the upstream region. Super-Alfvénic plasma flow is continuously injected from the left boundary ($x = 0$) with a speed of $\boldsymbol{V}_{in} = 3V_{A0}\hat{\boldsymbol{x}}$ (where $V_{A0}$ is the Alfvén speed in the upstream plasma). The particles are specularly reflected at the right boundary ($x = L_x$), forming a shock propagating in the $-x$ direction. Initially, a uniform background magnetic field is set with a large out-of-plane component: $\boldsymbol{B}_0 = B_0 \cos\theta_{Bn} \hat{\boldsymbol{x}} + B_0 \sin\theta_{Bn} \hat{\boldsymbol{z}}$, where $\theta_{Bn}$ is regarded as the shock normal angle. We adopt the proton-to-electron mass ratio of $m_p/m_e = 100$, the plasma beta $\beta_e = \beta_i = 0.01$, and the light speed $c = 100 V_{A0}$. Periodic condition is applied for both particles and fields in the $y$ direction. The spatial resolution is $\Delta x = \Delta y = 0.001 d_{i0}$, with each cell containing 100 computational particles initially. The time step is $5 \times 10^{-6} \Omega_{i0}^{-1}$ (where $\Omega_{i0} = eB_0/m_p$ is the ion cyclotron frequency). The electric field in the simulation is normalized by $E_0 = V_{A0} B_0$. The magnetic field and the number density of particles are normalized by their values in the upstream region ($B_0$ and $n_0$).

## 3. Results

## 3.1 Electron Pre-acceleration at the Shock Foot

The shock reaches a fully developed state after detaching from the reflecting wall, and it moves to the left with a speed of $\boldsymbol{V}_{sh} \approx -1.3 V_{A0} \hat{\boldsymbol{x}}$. Therefore, the Alfvén Mach number of the shock is $M_A \approx 4.3$ in the shock rest frame. Fluctuations in the electromagnetic field $\delta E_x$ and $\delta B_z$ at $t = 12\Omega_{i0}^{-1}$ are shown in Figures 1(a)-(b). The profile of the magnetic strength averaged over the *y*-axis is also inserted in Figure 1(b), displaying a typical quasi-perpendicular shock structure including the "foot", "ramp", and "overshoot". Electromagnetic fluctuations primarily propagating in the y direction is observed at the shock ramp (Figures 1(a)-(b), $106.0 < x/d_{i0} < 106.5$), which is known as the ripples of the shock surface. At the leading edge of the shock foot (region A marked by the red horizontal line, $104.1 < x/d_{i0} < 104.9$), the electric field exhibits short-wavelength fluctuations (Figure 1(a)). Since there are no magnetic fluctuations associated with them (Figure 1(b)), they are ES waves propagating in nearly the same direction.

Moreover, an enlarged view of the electron density fluctuations $\delta n_e$ in region A is shown in Figure 1(c), illustrating the same short-wavelength fluctuations. The formation of the shock foot is attributed to the reflection and gyration of upstream ions ahead of the shock ramp (Gosling & Thomsen 1985; Balogh & Treumann 2013; Yang et al. 2013). We identified these reflected ions and show their density distribution $n_{i,re}$ in Figure 1(d). These ions can be reflected as far as $x \approx 104 d_{i0}$ (Figure 1(d)), which matches the boundary position of the shock foot (Figure 1(b)). The coupling between the reflected ions and the electrons is obvious in region A (Figures 1 (c)-(d)), indicating the generation of ES waves due to the relative drift between incident electrons and gyro-reflected ions. This process, known as the ECDI, arises from the coupling of electron Bernstein waves with an ion beam mode carried by the reflected ions (Forslund et al. 1970). The Fourier-transformed wave spectrum of $\delta E_x$ in region A is shown in Figure 1(e). The ES waves mainly propagate in the -x ($k_x < 0$) and +y ($k_y > 0$) direction, which is consistent with the bulk velocity of reflected ions. The angle between $\boldsymbol{k}$ and the *x*-axis ranges from 16.6° to 57.5°, with a wavelength of $\lambda_{ES} = 2\pi/k \approx 0.045 d_{i0} \approx 9.86\lambda_D$ (where $\lambda_D$ is the Debye length in region A). Detailed properties of ECDI at the shock foot have been investigated in previous shock simulations (Muschietti et al. 2006, 2013; Yang et al. 2020). Here, we focus on the impact of ECDI on electron reflection and pre-acceleration.

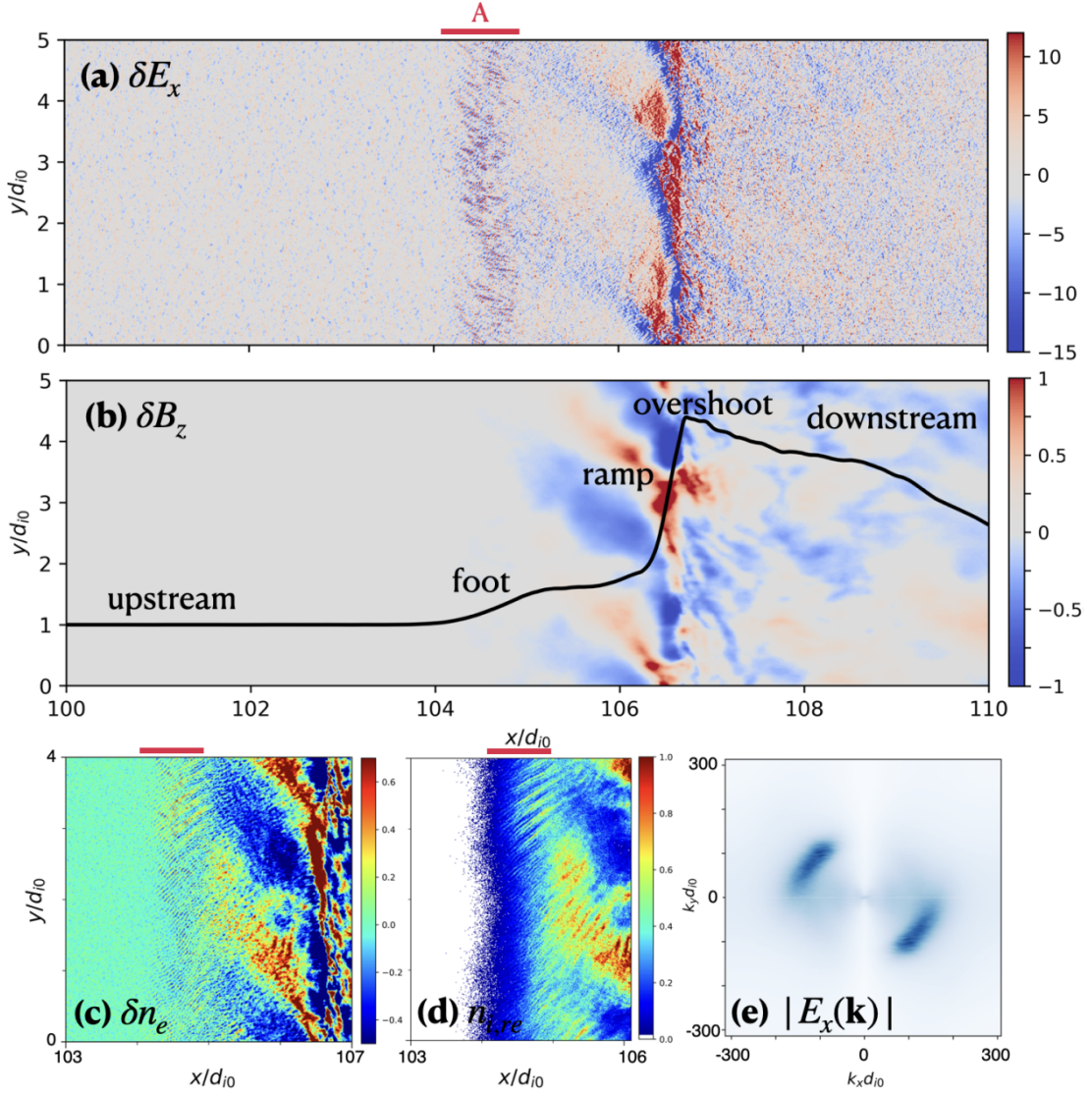

**Figure 1.** (a)-(b) The electromagnetic fluctuations $\delta E_x$ and $\delta B_z$ at $t = 12\Omega_{i0}^{-1}$. The fluctuations are calculated by subtracting the mean values along the *y*-axis from the local electromagnetic field (for example, $\delta E_x = E_x - \overline{E_x}$). The magnetic strength averaged along the *y*-axis $\bar{B}$ is also plotted by the black line in (b). (c)-(d) are enlarged views of the electron density fluctuation $\delta n_e$ and the density of the reflected ions $n_{i,re}$. The three red horizontal lines mark out region A ($104.1 < x/d_{i0} <$

104.9), where ES waves are obvious. (e) The Fourier-transformed wave spectrum of $\delta E_x$ in region A.

To analyze electron reflection by the shock, we switch to the de Hoffman-Teller (HT) frame (de Hoffmann & Teller 1950) for convenience. This frame is attained by introducing a velocity in the z direction at the shock rest frame so that the flow velocity is parallel to the magnetic field both upstream and downstream of the shock. Since the motional electric field vanishes in this frame, the reflection process of electrons can be considered as magnetic mirror reflection. Electrons will be reflected to the upstream if their pitch angles in the HT frame are greater than the loss-cone angle $\theta_{HT} = arctan(v_{HT,\perp}/v_{HT,\parallel}) > \theta_c = arcsin(\sqrt{B_0/B_m})$. Figure 2 shows the pitch angle distribution of the electrons at $t = 12\Omega_{i0}^{-1}$. The majority of upstream electrons ($x < 104 d_{i0}$) have pitch angles distributed between 0° to 15°, and only very few electrons can have pitch angles up to 45°. The maximum magnetic strength of the shock $B_m$ ranges between 3.6 to 4.5 after it is fully developed. We take the shock compression ratio of 4 ($B_m/B_0 = 4$), leading to $\theta_c \approx 30°$. According to this angle, only about 0.1% of the upstream electrons will be mirror-reflected by the shock. Apart from these background electrons, there is an electron beam moving antiparallel to the magnetic field in the upstream region. These electrons are reflected at the shock ramp ($x/d_{i0} = 106.8$) and move upstream, with their pitch angles ranging from 120° to 180°. However, the number density of these reflected electrons is about 1% of the upstream electron density, which is significantly larger than the number predicted by the mirror reflection model. This anomalous enhancement of reflected electrons is explained by microinstabilities in the shock foot region: The upstream electrons are perpendicularly heated and scattered to large pitch angles by ECDI before they encounter the shock ramp, making more of them get reflected. Significant scattering of their pitch angles mainly occurs at the leading edge of the shock foot (Figure 2, $104.1 < x/d_{i0} < 104.9$), consistent with the position of the ES waves induced by ECDI.

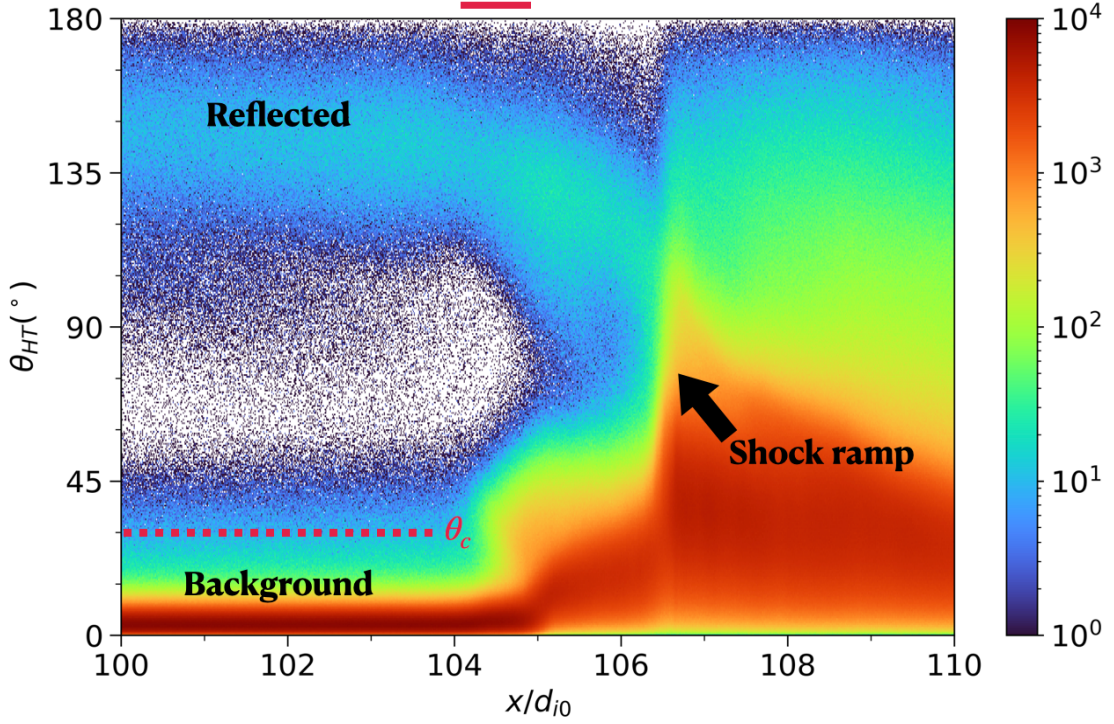

**Figure 2.** The pitch angle distribution of the electrons at $t = 12\Omega_{i0}^{-1}$ in the HT frame. The red dashed line marks out the loss cone angle of the shock. The red horizontal line above marks the position of region A.

To investigate the pitch angle scattering at the shock foot, we tracked 50074 electron particles located between $x = 105.5d_{i0}$ to $x = 105.6d_{i0}$ at $t = 10\Omega_{i0}^{-1}$. These electrons are positioned approximately $3.7d_{i0}$ upstream of the shock and travel downstream initially. We show the evolution of their position and velocity distribution in Figure 3. Additionally, we use the average magnetic strength acting on the tracked electrons ($\langle B \rangle$) to calculate the corresponding loss cone angle $\theta_{cp} = arcsin(\sqrt{\langle B \rangle/B_m})$ from $t = 10\Omega_{i0}^{-1}$ to $12\Omega_{i0}^{-1}$. At $t = 10.3\Omega_{i0}^{-1}$, the tracked electrons have not yet entered the shock foot and are about to encounter the ES wave region (Figure 3(a)). They exhibit a Maxwellian distribution in velocity space (Figure 3(e)), with only 0.1% of electrons displaying pitch angles larger than the loss cone angle. Subsequently, as the electrons approach the ES wave region, they undergo significant heating in the direction perpendicular to the magnetic field and are scattered to larger pitch angles (Figure 3(b) and (f)). This scattering process is approximately energy-conserving in the HT frame (Figure 3(e)-(f)). 3.2% of the tracked electrons have pitch angles larger than the loss cone angle after the scattering process, which is significantly higher than those at $t = 10.3\Omega_{i0}^{-1}$. The electrons are heated isotopically as they move further downstream (Figure 3(c) and (g)), but no significant scattering of pitch angles can be observed anymore (Figure 3(g)). There are still 3.2% of electrons that have pitch angles larger than the loss cone angle at $t = 11\Omega_{i0}^{-1}$. Eventually, most particles cross the shock ramp, while others with large pitch angles are reflected to the upstream region (Figure 3(d) and (h)). These results suggest that the heating by the ES waves at the leading edge of the shock foot ($t = 10.3\Omega_{i0}^{-1}$ to $10.6\Omega_{i0}^{-1}$) greatly increases the number of reflected electrons and provides sufficient energy for them to enter SDA.

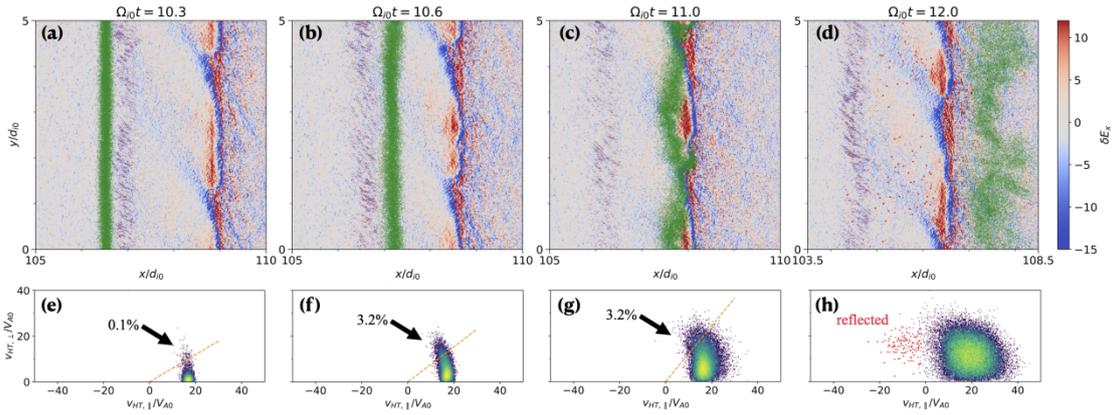

**Figure 3.** (a)-(d) Positions of the tracked electrons (green dots) overlaid on the electric field fluctuation $\delta E_x$. (e)-(h) Velocity distribution of the tracked electrons in the HT frame at the corresponding time. Orange dashed lines in (e)-(g) represent the loss cone angle calculated by $\theta_{cp} = arcsin(\sqrt{\langle B \rangle/B_m})$. Electrons that are eventually reflected to the upstream region are marked by red dots in (d) and (h).

We analyze the electron energization from upstream to downstream by decomposing the tracked electrons' mean energy gain $\Delta E$ into different parts in Figure 4. The perpendicular and parallel work is calculated as $W_\perp = -e \langle \int E_\perp v_\perp dt \rangle$ and $W_\parallel = -e \langle \int E_\parallel v_\parallel dt \rangle$. We take the integration time step as $\Delta t = 0.001\Omega_{i0}^{-1}$, which ensures that $\Delta E = W_\perp + W_\parallel$. As the shock is a slope of increasing magnetic strength (Figure 1(b)), electrons can be heated by adiabatic compression as they enter it. To distinguish the contribution of adiabatic compression and wave-particle interactions, we estimate the adiabatic heating as $W_{\perp,ad} = \langle \int [m_e v_{\perp,t}^2 (B_{t+dt}/B_t) - m_e v_{\perp,t}^2] dt \rangle$, where $v_\perp$ is evaluated in the electron fluid's rest frame at the position of each electron, and $B_t$ is the magnetic strength acting on the tracked electron at a given timestep. We also filter the electric field to eliminate fluctuations with small wavelength $\lambda < 0.06 d_{i0}$, so that we can distinguish the heating by large-scale structures and small-scale fluctuations. The work done by the filtered background electric field is calculated as $W_{\text{filtered}} = -e \langle \int \boldsymbol{E}_{\text{filtered}} \cdot \boldsymbol{v} dt \rangle$.

When the electrons are in the shock foot ($t = 10.3\Omega_{i0}^{-1}$ to $10.8\Omega_{i0}^{-1}$), $W_\perp$ is much larger than $W_\parallel$, which means $E_\perp$ acts as the main source of electron energization at the shock foot. A large part of electron heating at the shock foot comes from non-adiabatic perpendicular work, which occurs mainly between $t = 10.3\Omega_{i0}^{-1}$ to $10.6\Omega_{i0}^{-1}$ ($W_\perp - W_{\perp,ad}$ in Figure 4). This period corresponds to the same period in which the electrons pass through the ES wave region (Figure 3). The work done by the filtered background electric field is negligible in this region, which means the electron heating is mainly caused by small-scale ES waves located at the leading edge of the shock foot. These results suggest the upstream electrons are trapped by the ECDI-induced ES waves and non-adiabatically energized by the upstream motional electric field when they enter the shock foot.

As the electrons further move downstream, they encounter with the shock ramp at $t = 10.9\Omega_{i0}^{-1}$ to $11.3\Omega_{i0}^{-1}$. In this period, electrons are efficiently heated both in parallel and perpendicular directions. This heating process is highly non-adiabatic ($W_{\perp,ad} \sim 30\% \Delta E$). Electrons are mainly energized by large-scale structures in this region, as $W_{filtered}$ becomes dominant. These large-scale structures are related to the shock ripples and the shock potential electric field.

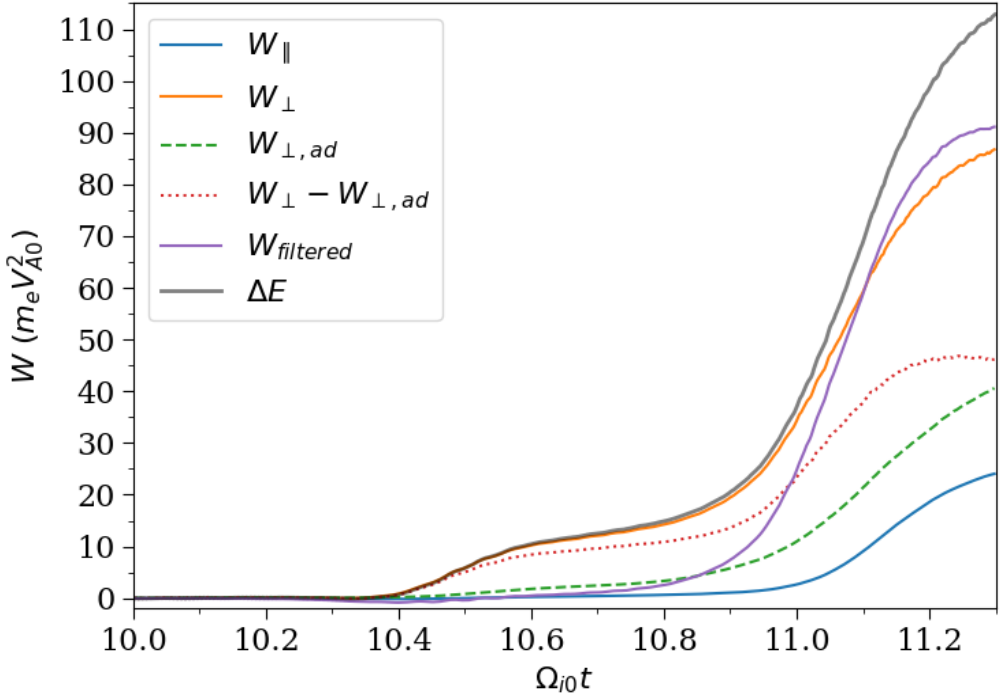

**Figure 4.** Mean work done on the tracked electrons from $t = 10\Omega_{i0}^{-1}$ to $11.3\Omega_{i0}^{-1}$.

Shock surface fluctuations observed at the shock ramp in Figures 1 and 3 may also affect the electron acceleration process. In order to compare the contribution of ECDI and shock ripples on electron reflection. We conduct another test-particle simulation. The electromagnetic field is extracted from the PIC simulation to push electrons at intervals of $0.001\Omega_{i0}^{-1}$. The gaussian filter is employed to filter out short-wavelength fluctuations of the electromagnetic field, so that the ES waves induced by ECDI are eliminated, but the large-scale shock surface fluctuations still exist (Figure 5(a)). The locations and the velocities of the test electrons are the same as the tracked electrons in Figure 3 at $t = 10\Omega_{i0}^{-1}$.

In contrast to the results in Figure 3, these test electrons are not scattered or heated when they enter the shock foot region between $t = 10.3\Omega_{i0}^{-1}$ to $10.6\Omega_{i0}^{-1}$ (Figure 5(a)-

(b)). The fraction of electrons out of the shock's loss cone remains below 0.1% untill they reach the shock ramp at $t = 11.0\Omega_{i0}^{-1}$ (Figure 5(e)-(g)). Very few electrons are reflected to upstream eventually at $t = 12.0\Omega_{i0}^{-1}$ (Figure 5(d) and (h)). Comparison between the results in Figure 3 and 5 shows that wave-particle interactions induced by ECDI are much more important than shock ripples in enhancing the electron reflection rate. Without the scattering by ECDI-induced ES waves, nearly no electrons can be reflected and accelerated by the shock.

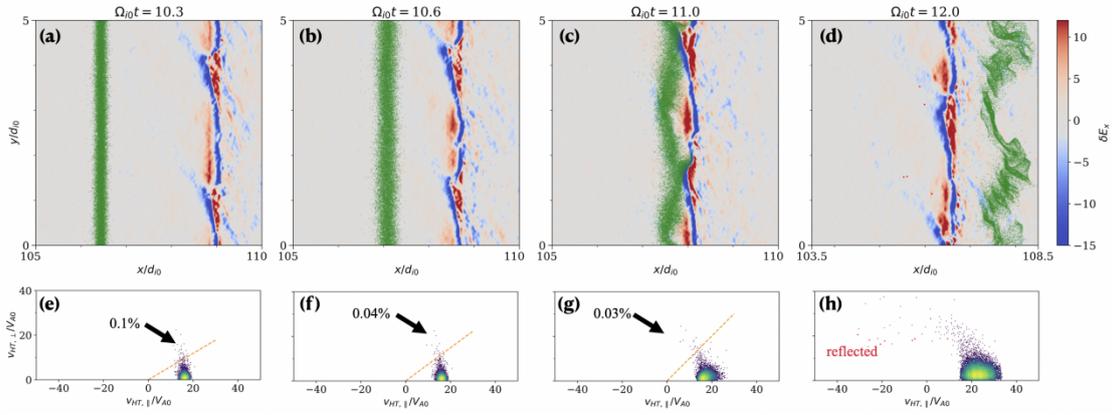

**Figure 5.** Electron tracking results in a test-particle simulation based on the filtered electronmagnetic field in the PIC simulation. (a)-(d) Positions of the tracked electrons (green dots) overlaid on the electric field fluctuation $\delta E_x$. (e)-(h) Velocity distribution of the tracked electrons in the HT frame at the corresponding time. Orange dashed lines in (e)-(g) represent the loss cone angle calculated by $\theta_{cp} = arcsin(\sqrt{\langle B \rangle / B_m})$. Electrons that are eventually reflected to the upstream region are marked by red dots in (d) and (h).

We further identify the shock surface by finding the first point satisfying $(B > 3.5B_0)$ in the x axis, which is shown by the grey lines in Figure 6. The shock surfaces appear to be structured in the y direction, rippling at a wavelength of about $3d_{i0}$. We pick out the reflected electrons upstream of the shock surface by defining their pitch angles in the HT frame larger than $80°$. The number density of the reflected electrons is found to be largely modulated by the shock ripples. At the bulges of the shock surface, more electrons are reflected than at other places (for example, at $x \sim 108 d_{i0}, y \sim 4.8 d_{i0}$ in Figure 6(a)). The average kinetic energy of these reflected electrons in every cell is also presented in Figure 6(b). It is observed that electrons reflected at the concaves of the shock surface are accelerated to higher energy (for example, at $x \sim 107.7 d_{i0}, y \sim 4 d_{i0}$ in Figure 6(b)), although their number density is lower in these positions. This phenomenon confirms that the shock ripples act as an extra scattering source, causing the electrons between two bulges to undergo stochastic acceleration while interacting with the ripples (Trotta & Burgess 2019). The results presented in Figure 5 and 6 suggest that shock ripples can not enhance the number of electrons who are able to enter the SDA process, but act as a modulation of electron reflection at different positions. Also, the acceleration efficiency of shock-reflected electrons can be enhanced by the shock ripples.

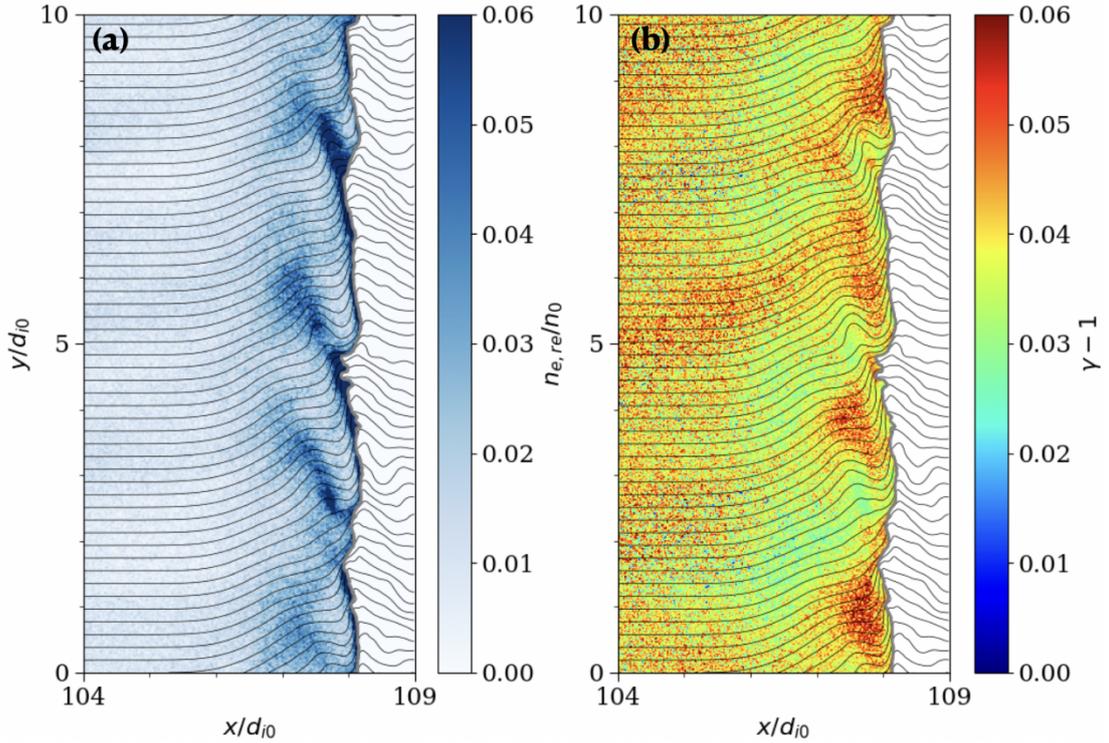

**Figure 6.** Shock surface fluctuations and their effects on electron acceleration at $t = 10\Omega_{i0}^{-1}$. The grey lines represent the identified shock surface. (a) the number density of the reflected electrons upstream of the shock surface. (b) the average kinetic energy of the reflected electrons in every computational cell.

## 3.2 In-plane Magnetic Field Configuration

The presented configuration with an out-of-plane magnetic field component may limit the possibilities of other kinetic instabilities, which can not be included due to their parallel propagation. To discuss this problem, we conduct another two-dimensional simulation for comparison. This case has the same parameters described in section 2, except that the upstream magnetic field is set in the x-y plane ($\boldsymbol{B}_0 = B_0 \cos\theta_{Bn}\,\widehat{\boldsymbol{x}} + B_0 \sin\theta_{Bn}\,\widehat{\boldsymbol{y}}$).

The structure of the shock at $t = 12\Omega_{i0}^{-1}$ is presented in Figure 7. Fluctuations of electron density and electric field are noticeable in the shock foot and ramp ($105 < x/d_{i0} < 107.5$). These fluctuations are mainly at ion kinetic scales, with their wavelengths ranging from 0.5 to $1 d_{i0}$. Unlike the out-of-plane configuration, short-wavelength waves at Debye scale are not observed in the shock foot. This is because the wave vector of the fastest-growing ECDI mode is parallel to the reflected ion beam. The growth of the waves and their effects on the electron acceleration are better resolved by the out-of-plane configuration, where the gyro-motion of the reflected ions is parallel to the plane of the simulation. The pitch angle distribution of electrons in the in-plane case is also presented in Figure 8. There is no shock-reflected electron beam moving anti-parallel to the magnetic field in the upstream region, indicating the fluctuations upstream of the shock front are not sufficient to scatter the electrons out of the shock's loss cone in the in-plane case.

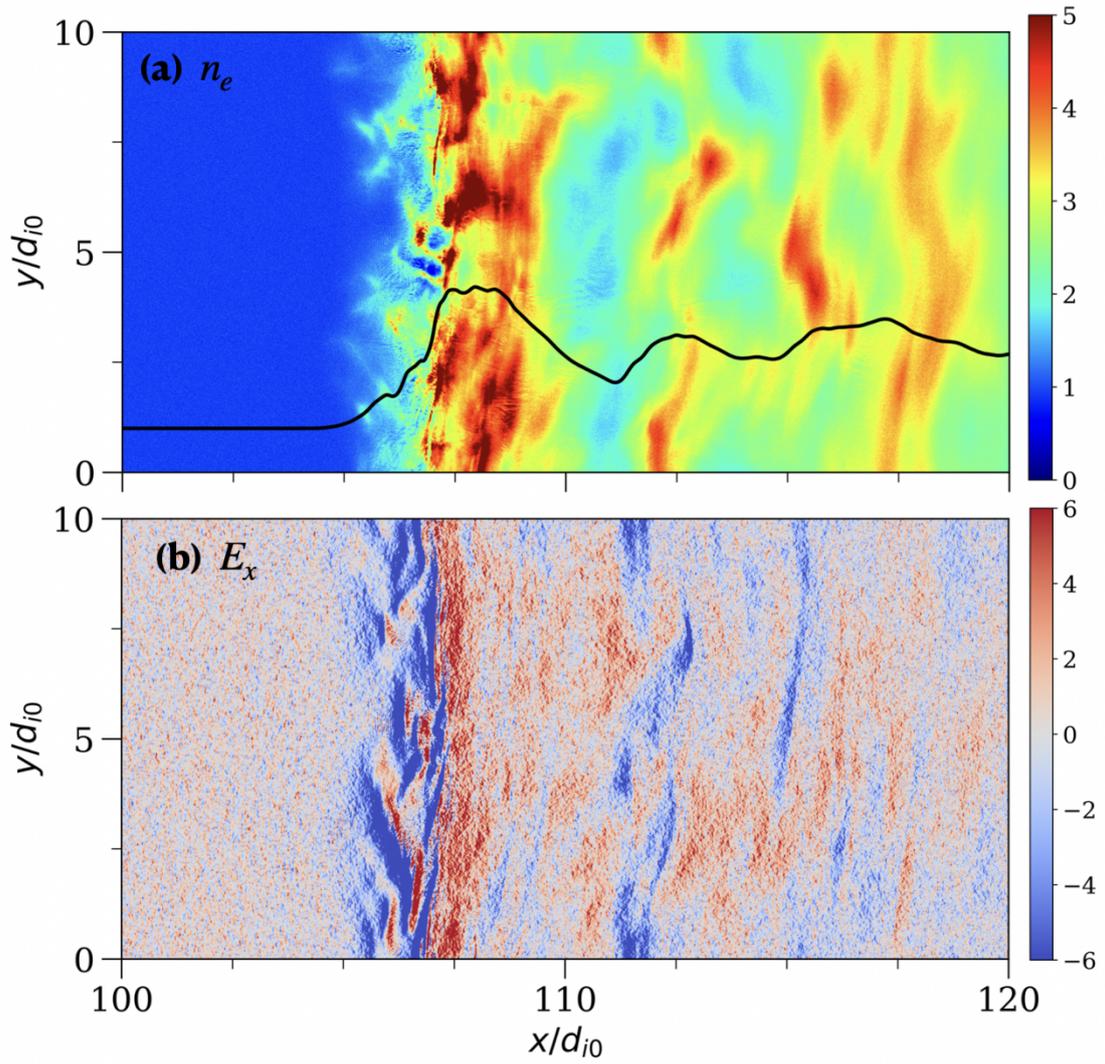

**Figure 7.** Shock structure for the in-plane magnetic field case at $t = 12\Omega_{i0}^{-1}$. (a) the number density of electrons. The magnetic strength averaged along the $y$-axis $\bar{B}$ is also plotted by the black line. (b) the electric field in the $x$ direction.

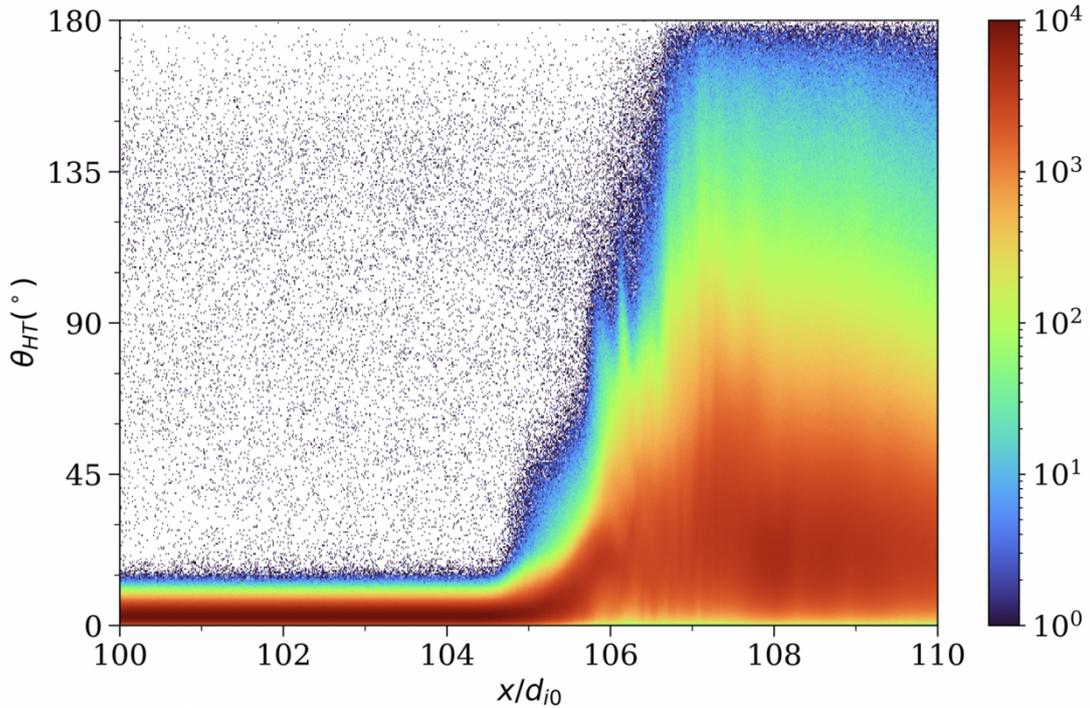

**Figure 8.** The pitch angle distribution of the electrons at $t = 12\Omega_{i0}^{-1}$ in the HT frame for the in-plane case.

## 4. Conclusions and Discussion

In conclusion, we have studied SDA at a low-Mach-number low-plasma-beta quasi-perpendicular shock. ES waves are found to be excited by ECDI at the leading edge of the shock foot, which can scatter the electrons into large pitch angles and make them more likely to be reflected at the shock ramp. Consequently, an unexpected number of electrons escape from the shock's loss cone and enter the SDA process. Besides, shock surface ripples are also found to affect electron acceleration, which modulates the electron reflection rates at different positions. These pre-acceleration processes pave the way for the injection of electrons into DSA at a low-plasma-beta quasi-perpendicular shock.

Observational evidences such as solar energetic particle events (Dresing et al. 2020, 2022) and type II solar radio bursts (Holman & Pesses 1983; Pulupa & Bale 2008) have shown efficient electron acceleration at CME-driven shocks, where the plasma beta and the Mach number are low (Maloney & Gallagher 2011; Poomvises et al. 2012; Morosan et al. 2019; Maguire et al. 2020; Bale et al. 2016; Shen et al. 2022). These phenomena can not be explained by the traditional mirror-reflection model, as it predicts insufficient reflection rate of electrons and low acceleration efficiency (Holman & Pesses 1983). Here, we find that interactions with microinstabilities can solve this problem. By scattering the electrons into large pitch angles, ECDI at the shock foot can significantly enhance the electron reflection rate, shedding light on the underlying mechanism at play in this challenging environment.

After entering the SDA process with the help of ECDI, the acceleration efficiency of shock-reflected electrons may be further enhanced by high-frequency wave activities in the shock ramp. Katou & Amano (2019) proposed a stochastic shock drift acceleration (SSDA) model that takes the effect of pitch angle scattering by high-frequency waves into the adiabatic SDA. In SSDA, electrons are confined at the shock ramp by stochastic pitch angle scattering off high-frequency waves, while gaining energy through SDA. This increases the interaction time with the shock ramp and hence provides more efficient acceleration than traditional SDA. Besides, as shown by observations and simulations, shock waves can generate a turbulent downstream region, which contains many magnetic islands (Gingell et al. 2019; Lu et al. 2021, 2024; Guo et al. 2023). Electron energization by magnetic reconnection between these magnetic islands may increase the subsequent acceleration efficiency (Zank et al. 2015). The transport properties of these superthermal electrons can be analyzed by self-consistent particle tracing, which helps to identify where the scattering occurs and how the energization takes place (Trotta et al. 2020).

To correctly resolve the gyro-motion of the shock-reflected ions and the excitation of ECDI, the out-of-plane magnetic field configuration is used in the main body of this article. Although the results in section 3.2 show that fluctuations in the in-plane case are insufficient to scatter or accelerate electrons, some parallel-propagating waves (such as the whistler mode waves) may be ignored in the out-of-plane case. Electron acceleration in three-dimensional shock simulations with compatible parameters would be an interesting topic for future research, which can self-consistently include both parallel and perpendicular fluctuations at the same time.

## Acknowledgements

This work was supported by the National Natural Science Foundation of China (NSFC) grants 42230201, 42174181, National Key Research and Development Program of China (No. 2022YFA1604600), and the Strategic Priority Research Program of Chinese Academy of Sciences, Grant No. XDB 41000000 and No. XDB0560000.